\documentclass[12pt]{article}
\usepackage[intlimits,tbtags]{amsmath}
\usepackage{amsfonts}
\begin{document}
%
\newcommand{\be}{\begin{equation}}
\newcommand{\ee}{\end{equation}}
\newcommand{\br}{\begin{eqnarray}}
\newcommand{\er}{\end{eqnarray}}
\newcommand{\lp}{\left(}
\newcommand{\rp}{\right)}
\newcommand{\lk}{\left\{}
\newcommand{\rk}{\right\}}
\newcommand{\lc}{\left[}
\newcommand{\rc}{\right]}
\newcommand{\sT}{{\scriptscriptstyle T}}
\newcommand{\2}{\,2}
\newcommand{\dif}{\mathrm{d}}
\def\a{\alpha}
\def\b{\beta}
\def\g{\gamma}
\newcommand{\se}{\section}
\newcommand{\tra}{\vec{p}_{\sT}}
\newcommand{\Z}{Z\left(\beta\right)}
\newcommand{\half}{\frac{1}{2}}
\newcommand{\ninteger}{\left.\mathbb{Z}\right.^n}
\newcommand{\nreal}{\left.\mathbb{R}\right.^n}
\numberwithin{equation}{section}

\thispagestyle{empty}

\begin{flushright}
\begin{tabular}{l}
FFUOV-02/10\\
{\tt hep-th/0211211}\\
\end{tabular}
\end{flushright}

\vspace*{2cm}

{\vbox{\centerline{{\Large{\bf Educing the volume out of the phase space boundary}}}}}

\vskip30pt

\centerline{Manuel A. Cobas, M.A.R. Osorio, Mar\'{\i}a Su\'arez
\footnote{E-mail addresses:
     cobas, osorio, maria@string1.ciencias.uniovi.es}}

\vskip6pt
\centerline{{\it Dpto. de F\'{\i}sica, Universidad de Oviedo}}
\centerline{{\it Avda. Calvo Sotelo 18}}
\centerline{{\it E-33007 Oviedo, Asturias, Spain}}

\vskip .5in

\begin{center}
{\bf Abstract}
\end{center}


We explicitly show that, in a system with T-duality symmetry, the
configuration space volume degrees of freedom may hide on the surface
boundary of the region of accessible states with energy lower than a fixed
value. This means that, when taking the decompactification limit (big volume
limit), a number of accessible states proportional to the volume is
recovered even if no volume dependence appears when energy is high enough.
All this behavior is contained in the exact way of computing sums by making
integrals. We will also show how the  decompactification limit for the gas of
strings can be defined from a microcanonical description at finite volume.

\vspace*{32pt}

\newpage

\section{Introduction}

  The density of states or its integrated version measuring the number of
accessible sates for a single object (we will denote it as $\Gamma$ and
reserve $\Omega$ for the density of states) can be considered as the
building block to get a description of a system with a big number $N$ (of
the order of Avogadro's number) of such objects. It is precisely this
quantity which provides us with the criterion to know whether classical
Maxwell-Boltzmann (MB) counting can be applied to a system by inspecting
whether the number of accessible states for one object is much bigger than
the number of objects in the system.

With the density of states for a single object, it is immediate to compute
$q= \sum_{i (\mbox{\tiny {states}})}\,\exp{(-\beta \epsilon_i)}$ by converting
the sum into one over energy levels. With $q$, the partition function for the
system with $N$ objects can be calculated by using  quantum or classical (MB)
statistics.

To compute $\Gamma(E,V,1)$, where we explicitly note that the number of
objects is one, it is necessary to previously look microscopically at the
problem of the single object at hand in some configuration in which a finite
volume can be defined. In general, this volume will be big as compared with
some other length or with energy if we have a fundamental constant with
dimensions of length at hand. Traditionally, dealing with objects that can
be abstracted as point particles, this is achieved by studying the quantum
mechanical dynamics of the object in a box of volume $V=L^d$, where the
walls of the box are represented by infinite barriers of potential. However
our problem is one in which the generic single object is a string. We take
the string as fundamental in various senses, but particularly in a sense in
which everything is made of or comes from strings. There is an immediate
definition of volume if we take our string and put it by hand in a space
which is compact ab initio, for example, if the space is a hypertorus of
volume
$V=(2\pi R)^d$ where $R$ is the value of the radius for every spatial
dimension as a simple election. Volume is then defined as a property of the
target space in which the string propagates. This is the kind of box we are
using here.

Strings are extended objects. As such, they have some kind of non local
features. What happens  to them in compact spaces when holes are present is
that winding is possible. Winding allows strings to know dimensions as a
whole. But, quantum mechanically, the wave function has also a kind of
non-local character that allows even a free particle to know that there
is a circular dimension and there is also a winding here measuring the number
of times the center of mass (one point) winds around the circle. The string
can simultaneously wind in the two fashions  so that  we
cannot distinguish one kind of winding from the other and finally we cannot
discern a big circle from a small one because each type of winding
respectively contributes to the energy in a way which is directly and
inversely proportional to the radius.

Historically, the computations of $\Omega(E,V,1)$ have been indirect by
using canonical (really macrocanonical with chemical potential $\mu=0$)
calculations to get this quantity from the partition function for a single
string by inverse Laplace transformation on the inverse temperature
variable. Final results have been gotten as an asymptotic approximation
valid for high energy (cf., for instance, \cite{salskager} and
\cite{tan}). To compute $\Omega (E,V,1)$ one has to sum over winding and
momentum number, when the approximation of substituting the sum over them by
a multiple integral holds, the output is a series over the energy
independent of the volume. From the sum itself it is also easy to see that
there is a regime which is proportional to the volume. Then, two important
questions emerge: for the single density of states, can we connect the high
energy regime independent of the volume and the one which is proportional to
it? For the gas of many strings, is there a well defined relation among
energy and volume for which one can say that the system behaves as the gas
of strings in uncompactified space? We will see that these two questions are
actually entangled and closely related to the way in which the system passes
from the low energy regime to the high energy situation as explained in
\cite{NH}.

The starting point of the analysis will be the simple form of $\Gamma$ as
a counting of states with energy lower than a prescribed generic value.
$\Gamma (E, R, 1)$ will be computed in section 2. 
There, we will explain the easy way in which the decompactification limit
can be defined and obtained from a simple relation between $R$ and $E$ and
how it can be connected to the description of the well known particle gas.
We will also explain the difficulties of applying an analogous reasoning to
try to get the dependence on $R$ (or the volume) from the sum over windings
and momenta when there is no such simple relation between energy and volume.
Furthermore, we will also justify the need to quantitatively know more about
the approximation of sums by integrals thorough the use of the
Euler-Maclaurin (EM) formula. So, section 3 we will be devoted to explain
what is relevant about that celebrated formula. In section 4, there appear
applications with special emphasis in showing up the way the volume
independent situation can be corrected thanks to the EM formula in order to
finally get the regime in which $\Gamma (E, R, 1)$ is proportional to the
volume. In section 5, we will present and explain the definition of the
decompactification regime for the gas with an Avogadro's number of strings
clearly showing that a decompactification limit can be defined upon the
microcanonical description of the system. Finally, section 6 will present
some comments and possible future prospects for the subject.

\section{Exact forms of $\Gamma$}

Our departure point will be the formulae giving the characteristic
perturbative spectrum of a closed string in a toroidal target space, namely

\br 
\alpha' E^{\2} = \frac{R^{\2}}{\alpha'}\vec{m}^{\2} +
\frac{\alpha'}{R^{\2}}\vec{n}^{\2} +
2N + 2\tilde{N}  \\
N - \tilde{N} + \vec{m} \cdot \vec{n}=0  \label{constraint}
\label{dispersion}
\er
 As usual, $\tilde{N}$ and $N$ stand  for the right and left moving
oscillator numbers and start from $-1$ if we deal with the closed bosonic
string and from zero for the closed superstring. So $\vec{m}$ and $\vec{n}$
are twenty five dimensional integer vectors in the bosonic case and nine
dimensional integer ones when supersymmetry is present. The second equation
is the left-right level matching condition, which now, in the compact case,
whenever finite winding and momentum are present, allows different right and
left moving oscillator numbers.

With this, it is easy to write $\Gamma(E,R,1)$ which represents the number
of states accessible to one string with energy lower than $E$. In a
hypertorus with $d$ (the number of spatial dimensions, i.e. $d=9
\,\mbox{or}\, 25$) circles with a common radius $R$ related to the volume by
$R=V^{\,1/d}/(2 \pi)$, $\Gamma$ reads
\be
\sum_{N,\tilde{N}} a(N,\tilde{N})\,\sum_{\vec{n},\vec{m}}\,\,
\delta_{N,\,\tilde{N}-\vec{m}\cdot\vec{n}}\,\,\vartheta \lp E^{\2}-
 \frac{R^{\2}}{\alpha'^{\2}}\vec{m}^{\2} - \frac{1}{R^{\2}}\vec{n}^{\2} - 
\frac{1}{\a'}\lp 2N
+ 2\tilde{N}\rp\rp
\label{anchura}
\ee

Here $a(N,\tilde{N}) = b_N \, b_{\tilde{N}}$ where the $b_N$ are well
known natural numbers. They represent the exponentially increasing degeneracy
of the oscillations of the string and are the origin of the Hagedorn
temperature in the canonical single string partition function
$q(\b)$. $\vartheta(x)$ is the Heaviside function whose value is $1$ if
$x>0$, zero if $x<0$ and $1/2$ when $x=0$. As expressed in \eqref{anchura},
it is easy to see that when
$R$ is big, $\Gamma(E,R,1)$ converts into a multiple sum in which only the
$\vec{m}=\vec{0}$ term (null winding) has a non vanishing contribution
because the Heaviside function in \eqref{anchura}
vanishes (for $\vec{m}\neq \vec{0}$) when $E < R/\alpha'$.

So, we have
\be
 \Gamma(E,R,1) =
\sum_{N} b_N^{\2}\,\sum_{\vec{n}}\,\,
\,\,\vartheta \lp E^{\2}-
 \frac{1}{R^{\2}}\vec{n}^{\2} - \frac{1}{\a'}4N\rp\,\,\,\,\,\lp
E<R/\alpha'\rp
 \label{clasica}
\ee

This can be seen as $\Gamma$ computed for an infinite number of relativistic
particles with squared masses $4N/\a'$. There is a total polarization
degeneracy of value $b_N^{\2}$ and the momenta are fully discrete. It is
also noticeable that the left-right level matching condition (a fully
closed-stringy ingredient) has become as trivial as in the open space case.
Anyway there still remains a stringy ingredient in the collection of
particles as a whole that is reflected in the exponential growth of the
coefficients $b_N$. This is the string seen as a collection of quantum
fields or analog model.

Now it is direct to use the standard argument to compute the sum as an
integral based on the fact that counting the number of accessible states
corresponds geometrically to compute the number of points of a cubic lattice
inside a sphere defined by the argument of the Heaviside function in
\eqref{clasica}. This is just accomplished by integration. This counting
approximates well the sum as long as the radius of the sphere is big enough
to contain a big number of points so as to minimize the relative error that
appears because of the clear bad counting near the spherical surface. We
will see in section 4 a precise way of computing this error. The result
of counting points by making an integral will finally give a term
proportional to the volume $V$.

It seems common sense to apply the same reasoning to  the crude
$\Gamma(E,R,1)$. However, there appear difficulties. Phase space geometry is
slightly more complex but, at first sight, tractable as the volume enclosed
by a kind of an elliptical deformation of an eighteen-dimensional sphere (to
be concrete, we focus on the superstring) producing two groups of nine
semi-axes each of squared length $ (E^{\2}-\frac{1}{\a'}(2N+2\tilde{N}))
R^{\2}$ and the T-duality transformed (i.e, what is obtained under
$R\longrightarrow \a'/R$) respectively. What spoils the reasoning of
counting states by computing volumes by integration is the closed string
level matching condition that makes the lattice one that cannot be built
upon replicating a cubic cell. However, with respect to the problem of
knowing how the volume dependence works in relation with computing
integrals, the fictitious system obtained by putting by hand $N=\tilde{N}$
instead of the constraint in
\eqref{constraint} will serve. The reasoning on counting states by
integrating to get a volume is then immediate and we may try to forget for a
moment what the error linked to surface effects could be, although we
anticipate it will be crucial. It is clearly easy to realize that the
phase-space volume of the elliptical object with the given axes is
independent of $R$ because, just with a simple change of variables with unit
Jacobian, we make $R$ disappear from the argument of the Heaviside function.
So, surprisingly enough, when the integral is a good approximation to the
sum, no volume dependence appears. Now, if we put this result and the
computation of the big volume limit (which gives the same result for the
closed string with constraint and the fictitious trivialized system)
together, it is clear that the difference between the sum and the integral,
i.e. what could be called the rest from a point of view approximating sums
by integrals, must be where volume dependence lies. It is then mandatory to
study the exact relation between computing sums by means of integrals and
sums themselves. This will be equivalent for us to studying the
Euler-Maclaurin summation formula in a multivariate form.

\section{The Euler-Maclaurin formula to make sums by means of integrals}

       Euler-Maclaurin summation formula was conceived and has mainly been
used as a celebrated formula for numerical integration by converting
integrals into sums. This formula was discovered independently by Euler and
Maclaurin around 1740. It can immediately be seen to be an extension of the
trapezoidal rule for integral approximation of one single variable
functions. We need it here to make the converse interpretation; i.e., to
compute sums by calculating integrals. Consequently, we need a multivariate
generalization.

Even having found a reference to the existence of such a multivariate
generalization in \cite{rangarao}, we have not been able to find the
equation giving it. So, in the appendix, we present a generalization based
on a systematic application of the Gauss theorem to a natural extension of
the beautiful proof for this formula presented in \cite{lerma1}(see also
\cite{lerma2} and the appendix here) based on the periodic  Bernoulli functions
of one single variable. The main idea roots on looking at the problem in a
distributional sense\footnote{Although it can also be understood only with
the need of the Stieltjes integration as in \cite{lerma2}.} and using the
periodic Dirac delta function as the natural mediator to convert sums into
integrals and vice versa. The Euler-Maclaurin summation formula will emerge
after realizing that the periodic Dirac delta function in $n$-dimensions
will produce one of the distributional generalizations of the Bernoulli
periodic functions. Being more explicit,

\be
\delta_{\mathrm{per}}(\vec{x}) = \sum_{\vec{k}\in\ninteger}\,
\delta(\vec{x}-\vec{k})
\label{periodic}
\ee
in such way that

\be
\label{withg}\begin{split}
\int_{\nreal}\,\,\dif\vec{x}\,g\lp\vec{x}\rp\,\vartheta\lp
f\lp\vec{x}\rp\rp\delta_{per}(\vec { x } ) &= 
\sum_{\vec{k}\in\ninteger}\,
\int_{\nreal}\,\,\dif\vec{x}\,g\lp\vec{x}\rp\,\vartheta\lp
f\lp\vec{x}\rp\rp\delta(\vec{x}-\vec{k})\\&= \sum_{\vec{k} \in
D}\mbox{}^{'}\,g(\vec{k}) = \sum_{\vec{k}\in \left.\mathbb{Z}\right.^n}\,
g(\vec{k})\,
\vartheta\lp f(\vec{k})\rp
\end{split}
\ee
where $\sum^{'}$ is a summation modified by taking only half of
$g(\vec{k})$ when $\vec{k}\in \partial D$, the surface 
defined by the equation $f(\vec{x})=0$ which bounds the compact region $D$
as a subset of $\mathbb{R}^n$. The one half factor comes from the definition
of the Heaviside function of one variable given above. It is worth to stress
that, when a product of more than one Heaviside function is necessary to
define the domain $D$, the prime will mean a certain combination of one half
terms for lattice points on the boundary (see the appendix). Now things
start to produce a multivariate Euler-Maclaurin formula because

\be 
1- B_0(\vec{x})=
\delta_{\mathrm{per}}(\vec{x})
\ee

Here $B_0(\vec{x})$ is obtained from $\vec{B}_1(\vec{x})$ from the relation

\be
B_0(\vec{x})= \vec{\nabla} \cdot \vec{B}_1(\vec{x})
\ee
where

\be
\vec{B}_1(\vec{x})= -\frac{1}{2\pi
i}\,\sum_{\vec{k}\in \lp\left.\mathbb{Z}\right.^{n}\rp^*}\,\,
\frac{e^{2\pi i \vec{k}\cdot
\vec{x}}}{\vec{k}^{\2}}\,\vec{k} =
-\frac{1}{2\,\pi}\,
\sum_{\vec{k}\in\lp\ninteger\rp^*}
\,\frac{\sin{(2 \pi \vec{k}\cdot\vec{x})}}{\vec{k}^{\,2}}\,\vec{k}
\label{B1}
\ee
which is an expression of $\vec{B}_1(\vec{x})$ thorough its Fourier expansion
that emphasizes its real character.

An exact formula can now be written in which it is explicit how a multiple
sum can be computed thorough integrals, to wit

\be
\left.\sum_{\vec{k}\in
D}\right.^{'}\,g(\vec{k}) = 
\int_{D}\dif\vec{x}\,\,g(\vec{x}) 
-\int_{\partial D}\,\, \dif\vec{S}
\cdot \lp g\lp\vec{x}\rp\vec{B}_1\rp +
\int_{D}\,\,\dif\vec{x}\,(\vec{\nabla}g)\cdot \vec{B}_1
\ee
Where we have used that

\be
\int_{\nreal}\,\,\dif\vec{x}\,
\vec{B}_1\cdot(\vec{\nabla}f)\,\delta\lp f\lp\vec{x}\rp\rp\,
g\lp\vec{x}\rp = - \int_{\partial D}\,\, d\vec{S}\cdot 
\lp g\lp\vec{x}\rp \vec{B}_1\rp
\label{surface}
\ee
with $\partial D$ the boundary of the region enclosed  by the surface
defined by  $f\lp\vec{x}\rp=0$ with a unit vector  normal to the surface
given by $\vec{e}_{\bf s} = - (\vec{\nabla}
f)/\lvert\vec{\nabla}f\rvert$. By simply taking
$\vec{B}_1= -\vec{\nabla}f/\lvert\vec{\nabla}\,f\rvert$ and $g(\vec{x})=1$,
\eqref{surface} is also useful to get $\lvert\dif\vec{S}\rvert$ when the volume
element is known because it is valid for any $\vec{B}_1$ and $g(\vec{x})$.

In the appendix we will show how to obtain more terms by using the other
multivariate generalizations of the Bernoulli functions given by
$B_{n\,\mathrm{(even)}}$ and $\vec{B}_{n\,\mathrm{(odd)}}$.\footnote{If,
after several applications of the Gauss theorem, the integral over $D$
representing the rest is suppressed, one finds that the remaining terms
generate an asymptotic series.}

\section{Simple applications to get $\Gamma (E, R, 1)$ for particles
and strings}

It seems justified to start this section satisfying our curiosity by the
simple application of Euler-Maclaurin's formula to the relativistic
classical gas. We suppose we have a single particle (of mass $M$) with
accessible energy levels
$E^{\2}=M^{\2}+\vec{n}^{\2}/R^{\2}$ in $d$ spatial dimensions. The lattice of
momenta is the one generated by the integers, i.e. $\mathbb{Z}^{\,d}$. The
equation defining the domain $D$ can be taken as $f(\vec{x})=(E^{\2} -
M^{\2})R^{\2} - x^{\2}=0$ where $x=\lvert\vec{x}\rvert$. It gives a sphere in
phase space of radius
$\rho=R(E^{\2}-M^{\2})^{1/2}$. The vector $\vec{e}_{\bf s}$ is the unit
radial vector. With all this we have

\be
\begin{split}
\sum_{\vec{n}}\vartheta(E^{\2}& - M^{\2} - \vec{n}^{\2}/R^{\2})  =  
\frac{2\pi^{d/2}}{d\,\,\Gamma(d/2)}\,\rho^{\,d} +
\,\rho^{d/2}
\sum_{\vec{k}\neq\vec{0}}\,k^{-d/2}\,\mathrm{J}_{d/2}\,\lp
2\,\pi\,k\,\rho\rp\\
 \xrightarrow{\,\rho\longrightarrow +\infty\,}&
\,\frac{2\pi^{d/2}}{d\,\,\Gamma(d/2)}\,\rho^{\,d}+
\frac{1}{\pi}\,\rho^{(d-1)/2}\,
\sum_{\vec{k}\neq\vec{0}}\,k^{-(d+1)/2}\,\cos\lp 2\,\pi\,k\,\rho - 
\lp\frac{\pi(d+1)}{4}\rp\rp
\end{split}
\ee

We see that, as expected, the absolute error coming from the surface term grows
with the radius of the sphere, i.e. grows with energy and $R$, but the
relative error of the surface contribution to the phase space volume of the
sphere decreases with energy and $R$. In terms of the configuration space
volume $V$, we see that, when $\rho$ is big, the relative error goes as
$V^{\,-(d+1)/(2d)}\xrightarrow{\mathrm{big}\,d}\,1/\sqrt{V}$. We remark that
the phase space surface term (what would be the error when approximating
sums by integrals) oscillates changing its sign. It is important to stress
that, for a single object with several discrete mass levels, whenever a mass
channel opens, there is a relevant contribution from the surface corrections
because $\rho < 1$ around that mass. The contribution just vanishes when the
channel opens.

The next application will be to the computation of the number of accessible
states for a single string for the fictitious trivialized system of section
2 with $N=\tilde{N}$ and unrestricted winding and momentum vectors. In this case
$g(\vec{x})=1$ and consequently $\vec{\nabla}g=\vec{0}$. The domain $D$ is
the region bounded by the surface given by
\be
1=\frac{\vec{m}^{\2}}{\frac{\alpha'^{\2}}{R^{\2}}\lp E^{\2}-4N/\a{\,'}\rp} +
 \frac{\vec{n}^{\2}}{R^{\2}\lp E^{\2}-4N/\a{\,'}\rp}
\label{ellip}
\ee
Here we keep $N$ fixed because we are interested in studying the dependence
on the volume $V$ by summing first on windings and momenta. Since
$g(\vec{x})$ is constant, the multivariate Euler-Maclaurin formula just
gives the difference between the sum and the integral in terms of the first
Bernoulli function as

\be
\sideset{}{'}\sum_{(\vec{m}\,,\,\vec{n})\in
D}\,1 -
\int_{D}\dif\vec{m}\,\dif\vec{n} =
-\int_{\partial D}\,\, \dif\vec{S} \cdot\vec{B}_1 
\ee

The integral over $D$ is the volume of the ellipsoidal region in the
18-dimensional phase space of momenta and windings that, as stressed in the
introduction, is clearly independent of the radius of compactification.

Our challenge will be to show (rather to prove out) that, when taking the
limit of big volume, the surface term, that gives the error made when
substituting the sum by the integral, is able to produce a contribution that
cancels the integral approximation and give also the decompactification
limit term that goes as $V$. We will take $d=9$ and focus on the superstring
case. After all, the bosonic string can only be used in this context as a
sort of a toy model.

To write the surface integral, we take polar spherical coordinates in both
the nine dimensional space of the variable $\vec{m}$ and the one for the
variable $\vec{n}$. Next, we take $n=\lvert\vec{n}\rvert= R
r\,\cos{\theta};\,\, m=\rvert\vec{m}\lvert= r (\a'/{R})\,\sin{\theta}$ with
$\theta \in [0,\pi/2]$. This turns \eqref{ellip} for the ellipsoidal surface into
the equation for a sphere of radius $(E^{\2}-4N/\a')^{1/2}$. The surface
element is given by
$$
\frac{\dif\,\vec{S}}{\dif\theta\,\,
\dif\Omega_{\bf m}\,\,\dif\Omega_{\bf n}} = 
\a'^{\,9}\lp E^{\2} - \frac{4N}{\a'}\rp^{17/2}\,
\sin^{\,8}{\theta}\,\cos^{\,8}{\theta}\sqrt{\frac{R^{\2}}{\a'^{\2}}\,
\sin^{\2}{\theta} + \frac{\cos^{\2}{\theta}}{R^{\2}}}
\,\,\vec{e}_{\bf s},
$$
 where 
$\Omega_{\bf n,\,\bf m}$
are the solid angles in the momentum and winding variables respectively. The
normal to the surface unit vector is given by

$$\vec{e}_{\bf s}=  
\lp \frac{R^{\2}}{\a'^{\2}}
\,\sin^{\2}{\theta} +
\frac{\cos^{\2}{\theta}}{R^{\2}}\rp^{-1/2}\,
\lp\vec{e}_{\bf m}\,\frac{R}{\a'} \sin{\theta}\,,\,
\vec{e}_{\bf n}\frac{\cos{\theta}}{R}\rp
$$
with $\vec{e}_{\bf m\,,\, n}$ the unit radial vectors for the space of the
continuous momentum and winding variables respectively.

We use \eqref{B1} to get $ \int_{\partial D}\,
\dif S\,\,
\vec{k}\cdot\vec{e}_{\bf s}\,\sin{(2\pi\vec{k}\cdot\vec{x})}/
\vec{k}^{\2}$ for a fixed mode 
$\vec{k}=(\vec{k}_{\bf m}\,,\,\vec{k}_{\bf n})$ and, after that,
sum over $\vec{k}\neq\vec{0}$. The surface integral equals
  
\be\begin{split}
\frac{1}{2\,\pi}\a'^{\,9}\rho^{17}\sum_{(\vec{k}_{\bf m}\,,\,\vec{k}_{\bf
n})\neq \vec{0}} 
\int&\,\dif\theta\,d\Omega_{\bf m}^{(8)}
\,\dif\Omega_{\bf n}^{(8)}\,\dif\theta_{\bf m\,,\,7}\,
\dif\theta_{{\bf n}\,,\,7}\,\,
\lp\sin{\theta_{{\bf m}\,,\,7}} 
\sin{\theta_{{\bf n}\,,\,7}}\rp^{\,7}\,\\ &\times\,
\lp\sin{\theta}\,
\cos{\theta}\rp^{\,8}\,
\frac{ k_{\bf m}\frac{R}{\a'}\sin{\theta}\cos{\theta_{{\bf m}\,,\,7}}
+ k_{\bf n}\frac{1}{R}\cos{\theta}\cos{\theta_{{\bf n}\,,\,7}}}{k_{\bf m}^{\2}
+ k_{\bf n}^{\2}}\\
&\times\,
\sin{\lp 2 \pi\rho\lp k_{\bf m}\frac{\a'}{R}\sin{\theta}\cos{\theta_{{\bf
m}\,,\,7}}
+ k_{\bf n}\, R\cos{\theta}\cos{\theta_{{\bf n}\,,\,7}}\rp\rp}\end{split}
\label{bigeq} 
\ee
where $\rho=(E^{\2}-4N/\a')^{1/2}$, and $\Omega_{\bf n\,,\,m}^{(8)}$ are the
solid angles for the eight-dimensional subspaces of the winding and momentum
variables orthogonal to the $X_{{\bf m}\,, 9}$ and $X_{{\bf n}\,, 9}$ axes
respectively. We have chosen the fixed vector $\vec{k}$ along a fixed
direction taking $\vec{k}_{\bf m}$ along the $X_{{\bf m}\,, 9}$ axis and
$\vec{k}_{\bf n}$ along $X_{{\bf n}\,, 9}$.

If we expand the sine function in the last factor of \eqref{bigeq} as
$\sin(2 \pi\rho k_{\bf m}\frac{\a'}{R}...)$
$\cos{(2 \pi \rho k_{\bf n}\, R...)} + \cos(..)\sin(...)$ we see that when
$k_{\bf n}\neq 0$ and $\rho\a'/{R}\ll 1$ the sinus function of the
$1/R$ term varies slowly as long as the cosine does rapidly producing a
small contribution to the integral. So, finally, in this limit
(formally,
$R\longrightarrow +\infty$) only the $k_{\bf n}= 0$ contributes
significantly and we can write for the integral

\be\begin{split}
\frac{R}{2\,\pi}\a'^{\,8}\rho^{17}\sum_{\vec{k}_{\bf m}\neq \vec{0}}& 
\int\,\dif\theta\,\dif\Omega_{\bf m}^{(8)}
\,\dif\Omega_{\bf n}^{(8)}\,\dif\theta_{\bf m\,,\,7}\,
\dif\theta_{{\bf n}\,,\,7}\,\lp\sin{\theta_{{\bf m}\,,\,7}} 
\sin{\theta_{{\bf n}\,,\,7}}\rp^{\,7}\,
\sin^{9}{\theta}\\&\qquad\times\,\cos^8{\theta}\,
\frac{\cos{\theta_{{\bf m}\,,\,7}}}
{k_{\bf m}}
\sin{\lp 2 \pi\rho\, k_{\bf m}\frac{\a'}{R}\sin{\theta}
\cos{\theta_{{\bf m}\,,\,7}}\rp}\end{split}
\label{biglimit}
\ee

The next step is to realize that whenever $\rho\a'/R\ll 1$ the sum over
$\vec{k}_{\bf m}\neq \vec{0}$ can be substituted by an integral. Since the
sum does not include the nine dimensional zero vector, we should exclude the
region in the lattice around the origin bounding it by a cube centered
in the origin. It is worth to note that one must be careful with the points
of the lattice that belong to the boundary of the cube because their
contribution will be corrected by powers of one half factors coming by the
vanishing arguments of the Heaviside functions defining the domain inside
the whole space. Schematically what we have is

\be 
\sum_{\vec{k}_{\bf m}\neq \vec{0}} =
\int_{\lp\mathbb{R}^{\,9} -\mathrm{cube}\rp}\quad +\quad 
\textrm{\small corrections for points that belong to the boundary}
\label{correc} 
\ee
In other words, the integral would approximate a sum of the primed type
previously described. To compute a sum without any restriction over the
points just on the boundary, we must correct the integral with the adequate
contribution for these points (see the appendix).  The integral taking out
the cube can be written as the difference between the integral to the whole
space minus the integral over the cubic region. This trick allows us to take
advantage of the spherical coordinates to compute for the whole space and
the Cartesian coordinates when calculating over the cube.

By making the sum over $\vec{k}_{\bf m}$ as an integral including $\vec{0}$
just after the other integrals are performed, we get that
\eqref{biglimit} equals the volume of a  nine dimensional sphere of radius
$R\,\rho$, i.e. $2 \pi^{9/2} (R\,\rho)^9/(9 \Gamma(9/2))$.
Now one has  to compute an integral over the region bounded by the  cubic
surface centered in the origin. In this compact region, $\lvert\vec{k}_{\bf
m}\rvert\leq \sqrt{9}$. This together with the fact that $R$ is big makes
$\sin{\lp 2 \pi\rho\, k_{\bf m}\,\a'\sin{\theta}
\cos{\theta_{{\bf m}\,,\,7}}/R\rp}\simeq 2 \pi\rho\, k_{\bf
m}\,\a'\sin{\theta}
\cos{\theta_{{\bf m}\,,\,7}}/R$ a good approximation as long as $R$ formally
goes to infinity.

The integral over the cubic region can then easily be done using Cartesian
coordinates. We have

\be\begin{split}
\a'^{\,9}\rho^{18}&\int_{-1}^{+1}\,\dif k_{{\bf m},1}...
\int_{-1}^{+1}\,\dif k_{{\bf m},9}
\int\,\dif\theta\,\dif\Omega_{\bf m}^{(8)}
\,\dif\Omega_{\bf n}^{(8)}\,\dif\theta_{\bf m\,,\,7}\,
\dif\theta_{{\bf n}\,,\,7}\,\lp\sin{\theta_{{\bf m}\,,\,7}}
\sin{\theta_{{\bf n}\,,\,7}}\rp^{\,7}\\
&\qquad\qquad\times\,\sin^{10}{\theta}\cos^8{\theta}\,
\cos^{\2}{\theta_{{\bf m}\,,\,7}}\,\,=\,\, \frac{2 \pi^{\,9} \rho^{18}}{18\,
\Gamma(9)}\,2^{\,9} \,\a'^{\,9}\end{split}
\ee
The presence of $\a'$ means that this term is of stringy nature. The factor
$2^{\,9}$ is the only contribution coming for the integrals partially
representing the sum over $\vec{k}_{\bf m}$. Now we need to correct this
result by adding the contribution of the points on the boundary. The rule is
clear (see the appendix): for the lattice vectors with a single component
equal to $\pm 1$ we have to add a term $1/2(f(1,0,...,0)+ f(-1,0,...,0)+
{\textrm{\small permutations}})$, for the lattice vectors on the boundary
which have two components equal to $\pm 1$ we have to add other term
$1/2^{\2}(f(1,1,0,...,0) + f(-1,1,0,...,0) + f(-1,-1,0,...,0)+
{\textrm{\small permutations}})$ and so on. The function $f$ is in fact a
constant because inside and over the cube the sinus function can be
approximated by its argument when $R$ is big enough. For each vector we then
have a constant contribution given by $2 \pi^{\,9}\,\a'^{\,9}
 \rho^{18}/(18\,\Gamma(9))$. The combinatorics involved is not too difficult.
The vectors that contribute $1/2$ of the constant value are the ones with
null components but one single $+1$ or $-1$; there are
$2\,\dbinom {9}{1}$ such vectors. The ones
which give $1/2^{\2}$ of the constant value are the ones with two components
equal to $\pm 1$; there are $2^{\2}\,\dbinom{9}{2}$ of them, and so
on. The total correction from all the lattice points on the cubic boundary
is

\be
\frac{2 \pi^{\,9} \rho^{18}}{18
\Gamma(9)}\,\,\a'^{\,\,9}
\,\sum_{i=1}^{9}\,\binom {9}{i}= \frac{2 \pi^{\,9} \rho^{18}}{18\,
\Gamma(9)}\,\,\a'^{\,\,9}\,\lp 2^{\,9} -1 \rp
\ee

Consequently, we have shown the way the volume reappears if one starts from
a regime in which the integral is a good approximation to the sum to compute
the number of accessible states for a single object. The whole calculus for
the trivialized system can be summarized as

\be
\begin{split}
\sum_{\vec{n},\vec{m}}
&\vartheta \lp E^{\2}-
 \frac{R^{\2}}{\alpha'^{\2}}\vec{m}^{\2} - \frac{1}{R^{\2}}\vec{n}^{\2} -
\frac{4\,N}{\a'}\rp = \\
& \int_{\mathbb{R}^9\times\mathbb{R}^9}d\vec{m}\,\dif\vec{n}\,
\,\,\,\vartheta \lp E^{\2}-                                              
 \frac{R^{\2}}{\alpha'^{\2}}\vec{m}^{\2} - \frac{1}{R^{\2}}\vec{n}^{\2} -
\frac{4\,N}{\a'}\rp - \int_{\textrm{Ellip.}} \dif\vec{S}
\cdot\vec{B}_1\\ & 
\xrightarrow{\,\,\,\,\, \rho\a'/R\,\ll\,1\,\,\,\,\,}\,\,\,\,
\frac{\a'^{\,9}\rho^{18}\,\pi^{\,9}}{362880} + 
\lc\frac{32\,\pi^4\,\lp\rho\,R\rp^{\,9}}{945} -
\frac{\a'^{\,9}\rho^{18}\,\pi^{\,9}}{362880}\lp\, 2^{\,9} - \lp
2^{\,9}-1\rp\rp\rc\end{split}
\ee

We are now prepared to confront the pure  string system. The complication
added to the trivialized system is the left-right level matching condition
that is a function of winding and momentum numbers. We have to find a
representation of the Kronecker delta in \eqref{dispersion} in a way that
can be used by the Euler-Maclaurin formula thorough a function of real
variable as $g(\vec{x})$ in \eqref{withg}. To this purpose, we represent the
Kronecker delta as an integral

\be
\label{ligadura}
\int_{-1/2}^{+1/2}\,\,\dif t\,\, \exp{(- 2\,\pi\,\mathrm{i}\,t\,(\Delta -
\vec{m}\cdot\vec{n}))}=\delta_{\Delta\,,\,\vec{m}\cdot\vec{n}}=
\frac{\sin(\pi\,(\Delta -\vec{m}\cdot\vec{n}))}{\pi\,(\Delta -
\vec{m}\cdot\vec{n})}
\ee 
which holds for $\Delta = \tilde{N} -N$  integer and $\vec{m}$, $\vec{n}$
integer vectors.

The number of accessible states for one string which has energy lower than
a fixed value $E$ can then be written as

\be
\begin{split}
\Gamma(E,R,1)= & \sum_{N,\,\tilde{N}}
a(N,\tilde{N})\,
\sum_{\vec{n},\vec{m}}\,\,\int_{-1/2}^{+1/2}\,\dif t\,\,
\mathrm{e}^{- 2\,\pi\,\mathrm{i}\,t\,\lp\Delta -
\vec{m}\cdot\vec{n}\rp}\\
 &\times\, \vartheta \lp E^{\2}-
 \frac{R^{\2}}{\alpha'^{\2}}\vec{m}^{\2} - \frac{1}{R^{\2}}\vec{n}^{\2} -
\frac{1}{\a'}\lp 4N
+ 2\Delta\rp\rp 
\label{anchura2}
\end{split}    
\ee

For given oscillator numbers, we can  represent the sums over windings and
momenta as integrals by using the multivariate Euler-Maclaurin formula to 
get 

\be 
\label{grande}
\begin{split}
&\int_{-1/2}^{+1/2} \dif t \,\mathrm{e}^{-2\,\pi\,\mathrm{i}\,t\,\Delta}
\sum_{\vec{n},\vec{m}}
\mathrm{e}^{2\,\pi\,\mathrm{i}\,t\, \vec{m}\cdot\vec{n}}\\
&\qquad\times\vartheta \lp E^{\2} -
 \frac{R^{\2}}{
\alpha'^{\2}}\vec{m}^{\2} - \frac{1}{R^{\2}}\vec{n}^{\2} -
\frac{1}{\a'}\lp 4N
+ 2\Delta\rp\rp\\
&\quad =
\a'^{\,9}\,\,E^{\,18}\,\int_{-1/2}^{+1/2} \dif t
\quad\mathrm{e}^{\,-2\,\pi\,\mathrm{i}\,t\,\Delta} 
\lc\quad\int_{D}\,\dif\vec{m}\,\dif\vec{n}\quad
\mathrm{e}^{\,2\,\pi\,\mathrm{i}\,t\,\a'\,E^{\2}\vec{m}\cdot\vec{n}}\right. \\ 
&\left.\qquad -\, 2\,\int_{\mathbb{R}^{\,9}\times\mathbb{R}^{\,9}}\,\, 
\dif\vec{m}\,\dif\vec{n}\,\,
\mathrm{e}^{\,2\,\pi\,\mathrm{i}\,t\,\a'\, E^{\2}\,
\vec{m}\cdot\vec{n}}\,
\lp\frac{R\vec{m}}{\a'}\,,\,\frac{\vec{n}}{R}\rp\cdot
\vec{B}_1\lp \frac{E\a'\vec{m}}{R}\,,\,
E\,R\,\vec{n}\rp
\right.\\
&\left.\qquad\qquad\qquad\times\,E^{\,-\,1}\,\delta\lp 1-\vec{m}^{\2} -
\vec{n}^{\2}- \frac{4N+2\Delta}{E^{\2}\,\a'}\rp
\right.\\
&\left.\qquad +\, 2 \pi\,\mathrm{i}\,t\,E\,
\int_{D}\,\,\dif\vec{m}\,\dif\vec{n}\,
\quad\mathrm{e}^{2\,\pi\,\mathrm{i}\,t\,\a'E^{\2}\,
\vec{m}\cdot\vec{n}} 
\lp R\,\vec{n}\,,\,\frac{\a'\vec{m}}{R}\rp
\cdot \vec{B}_1\lp\frac{E\a'\vec{m}}{R}\,,\,E\,R\,\vec{n}\rp\rc
\end{split}     
\ee 
where the domain $D$ is the region bounded by the surface defined by the
vanishing of the argument of the Dirac delta that appears because it seems
convenient to represent the surface integral as a volume integral over the
whole (continuous) phase space.

Now one has to compute the limit of $\Gamma$ when the volume gets big to see
the way it finally becomes proportional to the volume of the hypertorus as
one departs from a situation in which no volume dependence appears because
the first integral is a good approximation. When energy is high enough, one
gets that $D$, the domain of integration\footnote{$D$ is always an
ellipsoid with semi-axes given, after putting $\rho\,'=(E^{\2}-
(4N+2\Delta)/\a')^{\,1/2}$, by
$\rho\,'\a'/R$ and $\rho\,'R$}, is such that the number of points inside the
boundary is really big when compared to the error over the surface that
increases, but with a lower power of
$\rho'$.

 Let us now give the detailed proof of the way the high volume limit
shows up from \eqref{grande}. The first step is just to expand
$\mathrm{e}^{\,2\,\pi\,\mathrm{i}\,t\,\a'\,E^{\2}\vec{m}\cdot\vec{n}}$ as a
series of powers of its argument. The zero order in this expansion gives, for
the first and second integrals, the same contribution as for the fictitious
trivialized toy model (after changing $\rho$ by
$\rho\,'$), so the analysis made for this example applies for it and finds
here its real purpose. The contributions from the odd powers
vanish\footnote{We could have written
$\,\cos{\,2\,\pi\,\,t\,(\Delta -
\vec{m}\cdot\vec{n})}$ instead of the exponential as the integrand in
\eqref{ligadura} to make this manifest.}. For the last integral in
\eqref{grande},
it is easy to see that the contributions from the even powers of the
expansion of the exponential also vanish.

We have to compute a sum over $\vec{k}_{{\bf m}}$ in the integrals
containing $\vec{B}_1$. This sum can be computed as an integral over the
real variable $\vec{k}$ plus the contribution from taking out the point
$\vec{k}_{{\bf m}}=\vec{0}$ (see \eqref{correc}). The contribution
integrated over $\vec{k}\,\in\,\mathbb{R}^{\,9}$ coming
from the $(2j-1)$-th power in the third integral exactly cancels the part
integrated over $\vec{k}$ from the $2j$-th power of the second
integral (always in the limit in which $\rho\,^{'}\a'/R \ll 1$).

Finally, for both sums over $\vec{k}_{{\bf m}}$, we are left with the
contributions from the points on the boundary of the cube centered at the
origin plus the integral over this domain (cf. \eqref{correc}). It is easy
to see that these contributions from the $2j-1$-th order in the third
integral and those from the $2j$-th order in the second one add up to
exactly cancel the $2j$-th order of the first integral, which is the part
manifestly independent of the configuration volume
$V$. This is the rigorous proof of the connection between the particle gas
regime and the volume independent situation for $\Gamma (E, R, 1)$.

\section{The limit of decompactification for the gas of strings}
 
Now it is mandatory to try to get the first consequence of what we have
learned. We think that the issue on whether the uncompactified regime for
the gas of strings can be recovered from the compact description deserves a
certain reanalysis after putting together what we know now on $\Gamma (E,
R,1)$ and what is presented in \cite{NH}.

By using the inverse Laplace transform techniques the best
that has been obtained for the density of states of the gas of strings with
compact dimensions is a series that converges whenever \cite{tan}

\begin{equation}
E > K
R^{\,d}/{\alpha'}^{(d+1)/2}\equiv E_H
\label{EH}
\end{equation}
($d$ is the number of big dimensions). The term on the right hand side of
this inequality is a sort of Hagedorn characteristic energy. The important
point here is that in
\cite{NH} it is explained that, when $R$ is big enough as compared with
$\sqrt{\a'}$, $K$ can actually be computed to give a
precise physical definition of the Hagedorn energy $E_H$ as the energy at
which the system reaches the Hagedorn temperature or approaches it from a
low energy regime dominated by a gas of massless objects. Hence, the
Hagedorn regime will be defined as the regime at which the massive modes
appear in the gas. These modes have masses that are proportional to
$1/\sqrt{\a'}$ which is the characteristic mass scale of the stringy
effects. They are then oscillators, whose mass is a positive integer
times $1/\sqrt{\a'}$, or windings of mass $R/{\a'}$.
 To be more concrete, when one has a system with an Avogadro's number of
strings, looking at its Hagedorn regime dominated by oscillators means
looking at the system when the the energy is such that
$E/N_H > 2/\sqrt{\a'}$, where $N_H$ is the number of strings that the gas
presents at the Hagedorn energy as given by the characteristic Hagedorn
temperature (cf. \cite{NH}).

\begin{equation}
E_H=C_{10}\,V\,\beta_H^{-10}
\end{equation}
where $C_{10}$ is a known  constant.

 The bound $2/\sqrt{\a'}$ is precisely the minimum mass of a non vanishing
vibrational mode of a single string. At this point, we must take into
account that there is a relationship between the number of strings and
energy and also volume (or $R$). It is the result of the fact that our
microcanonical description is one very special (it is not rigorously
microcanonical) for which the chemical potential $\mu$ vanishes. This
condition for equilibrium gives in
\cite{NH} that $N_H \sim V/\a'^{d/2}$, which simply states that the number
of strings at this characteristic energy is finally proportional to the
volume, like in the black-body in equilibrium with a gas of photons. This is
so because one can always think of the system as reaching or approaching the
Hagedorn temperature $T_H$ at the energy
$E_H$ departing from the low energy regime in which strings are in their
massless modes. Thus, $E_H$ is the energy which separates the low energy
regime of the gas dominated by the massless modes of the string (the
$\a'\longrightarrow 0$ limit taken on the mass formula) from the string or
high energy regime in which vibrational or winding modes appear. If $R$ is
big enough, there is always a low energy regime with massless modes for the
gas so we now understand why the condition $E-E_H > 0$ appears the same for
different ab initio topologies. This was seen as something notorious in the
past.

If one naively thinks that the condition to obtain the uncompactified result
is to impose $E < R/\a'$, i.e., the replica of the condition over the
single string, but now $E$ stands for the total energy, it is easy to see
that being in the Hagedorn regime together with this condition forces
$R$ to be small instead of big as it was supposed. This way, it would seem
that there is no domain of energy and radius for which the characteristic
asymptotic behavior in uncompactified space can be recovered. The physical
picture of this behavior at the Hagedorn regime is that of a single string
absorbing all the energy in its vibrational modes and approaching
equilibrium with a sea of low energy strings at the microcanonical Hagedorn
temperature. It is precisely the physical picture what makes clear that, to
obtain the Hagedorn regime dominated by the vibrational modes characteristic
of the uncompactified result, the condition to add to that of being in the
Hagedorn regime ($E > E_H$) is
$E-E_H < R/\a'$. It simply states that the single fat string cannot have
winding modes. Now, this is compatible with the fact that
$R \gg \sqrt{\a'}$. In other words, to get the fat string dominated regime
characteristic of having open spatial dimensions, 
one needs a radius big enough so as to have that the energy of the system
minus the Hagedorn energy does not suffice to get windings in the gas in
equilibrium (or quasi-equilibrium). Vibrational modes are much lighter and, 
after reaching $E_H$, massive strings appear to stay in meta-equilibrium
with the sea of massless modes. It is worth to notice that the volume can be
very big but the momentum that a string in the gas can have still belongs to
a discrete set; we then have and effective decompactification but not a pure
one that would be the one  in which the nine torus certainly 
becomes $\mathbb{R}^{\,9}$ and the momenta get dense. We will treat this
point later when commenting on the Jeans instability problem.

What  happens with the approximate series computed in \cite{tan} is that it
is adapted to describe the Hagedorn regime when $E-E_H > R/\a'$.
Mathematically this is analogous to the way in which the integral
approximates the sum in $\Gamma$ to only describe the high energy
approximation.

If, from the very beginning, one has that $R$ is of the order
of $\sqrt{\a'}$, then the mass formula tells us that there is no radiation
dominated phase (the standard black-body regime with massless objects)
because a massless object with non vanishing momentum is indistinguishable
from a massless object with a non vanishing winding. However equation
\eqref{EH} is still valid with $d=0$ giving  that high energy means $E >
K /\sqrt{\a'}$ (cf. the way this condition appears in \cite{tan}). In any
case, the total energy has to be bigger than the minimum momentum or winding
mass which is as big as the mass of a vibrational mode. Of course, if
winding modes are present in the gas, from the expression approximating
$\Omega (E,R)$, it seems that one cannot get the characteristic negative
specific heat phase for the gas with uncompactified dimensions. But,
contrary to what is stated in \cite{tan} this
does not in any way imply that there is no domain in the plane
$(E,R)$ for which the uncompactified behavior can be found.

\section{Comments and outlook}

Before commenting on the contains of the work, may be some
more general remarks come into place to justify the interest of the topic 
we have dealt with. First, it is  important to notice that
perhaps one of the most active research topics in String Theory
has been the one about the 
Hagedorn phase which is characteristic of a gas of {\it free} and 
{\it perturbatively interacting} strings. 
Several published works have claimed to
have completely solved the problem mainly in two ways: there is really
a (may be first order) phase transition at the Hagedorn temperature or
the Hagedorn temperature is a maximum one. Things are more
subtle than some statements that have been published and 
are being taken as true. This is the motivation for us to pursue a
plan to unambiguously settle  what is really well established and what is
simply wrong when treating the topic. To this purpose we have already
clarified the way quantum statistics effects are
taken into account in the problem, the connection between the presence of
non extensive terms and negative microcanonical specific heats, and the
physical meaning and value of such peculiar systems \cite{NH}. 
Systems having a negative microcanonical 
specific heat are a subject of intense active research 
by the community of theoreticians working on Statistical Mechanics and by
atomic and condensed matter experimentalists. 

The main criticism that can be made to any work on the free string gas is very 
well known because
one can find it, for example, in a classic review on String
Theory  as it is the two volume work by M. Green, 
J. Schwarz and E. Witten \cite{GSW}. It is a criticism that simply brushes away 
the Hagedorn problem because Jeans instability and the nucleation 
of black-holes simply make ill-defined the canonical treatment when
interactions are plugged in. 
It is then often argued, when treating thermodynamical properties of string 
gases, that no thermodynamic limit can be taken because we are dealing with
systems that include gravity, i.e. long range interactions, and no infinite
volume (thermodynamic) limit exists (although there is certain 
controversy on this subject,
it seems that the theorem presented  in \cite{laliena1} 
is an exact
proof  of this fact for classical gravitational interactions). 
As we are unable to take this limit, the concept of
temperature and the idea of a possible phase transition or having a maximum
temperature must be discarded because we will be unable to follow the ideas
coming from Statistical Physics (see again \cite{GSW}).

This statement is both wrong and true; it needs clarification. When no
thermodynamic limit exists, the canonical description, whose main parameter
is temperature ($T$), becomes ill-defined. The reason is that, from the
mechanical point of view, $T$ is a derived parameter, which has no direct
significance. On the other hand, a microcanonical description relies upon
the use of energy $E$, a quantity conserved for an isolated system and well
defined in phase space. Entropy has also a clear meaning as the logarithm of
a volume in phase space (we do not really need a probabilistic definition:
$S= k \mathrm{ln} W$ versus $S= \sum_i\, p_i\mathrm{ln}\,p_i$). In
Statistical Thermodynamics, extensivity is assumed and the thermodynamical
limit (as $V\rightarrow \infty$) is taken. It is in this situation that
both canonical and microcanonical descriptions appear to be equivalent. A
phase transition will show as a non homogeneity of the systems, a non
extensivity.

In our problem, even when the string coupling vanishes ($g_s=0$) and we do
not have long range interactions, we do not have an extensive entropy
\cite{NH}. It can also be shown that microcanonical and canonical
descriptions are, in some cases, inequivalent.
After introducing interactions in our system, no thermodynamic limit can be
taken as the result of the anti-screening properties of gravitational
forces; we will then be forced to treat the system microcanonically, this being
the ensemble adapted for "small" systems. This would let us also study phase
transitions in finite volume systems which, by the way, can also be seen
experimentally in "small" systems. A very
clear exposition of these facts and its relation to negative specific heats
can be found in \cite{grosscondmat0004268} where, among others,
classical self-gravitating systems are specifically treated.

For the particular case of self-gravitating systems, where Jeans instability
appears, an important effort has been made to treat the problem. 
In the work by Gross-Perry-Yaffe \cite{gpy}, 
where an approach from the quantum field theory point of view was presented, 
the need for a microcanonical treatment was remarked. Since then, 
the traditional
reasoning  to avoid Jeans' instability is to put the system in a finite
volume as we have done in our work by using a hypertorus. Then the 
procedure is to rapidly scale the
string coupling constant to zero keeping  the canonical energy
density constant so as to allow the volume to become large. Alternatively, 
one can fix the coupling as very small and restrict 
the volume to be large but finite\footnote{More concretely one has that
$\overline{E} < R^{\,d-2}/(g_s^{\2}\alpha'^{(d-1)/2}) \sim R^{\,d-2}/G_N$ 
as the condition to avoid Jeans' instability}. 
It is standard lore that to solve
infrared divergences due to long range interactions one can use a finite
volume. With finite volume there is no $p^{\,\mu}=0$ momentum for the massless
objects. Jeans' instability really appears thorough the computation of a non
vanishing contribution for  the temporal components of the
vacuum polarization tensor with $p^0=0$ and $\vec{p}\rightarrow \vec{0}$ 
in the pure decompactification limit in which the momenta become dense, in
particular, around zero. 
To avoid  Jeans' instability in the canonical 
description, one can have a really big
radius (volume) as compared to $\sqrt{\alpha'}$, but the momenta must still
be discrete. This is the precise meaning of 'big but finite' and the difference
between an effective decompactification (big $R$) and a pure
decompactification in which the torus becomes $\mathbb{R}^{\,9}$ and the
momenta become then dense.
  
A very remarkable  fact is that,
with an up to now complete ignorance from the string theorist community, 
the treatment of self-gravitating systems  
has recently been a very active research field
in Statistical Physics \cite{lalienayotros}. Although, in
these works, gravity is treated classically (as Jeans did), very interesting
results appear related with phase transitions for the system 
stemming from negative specific
heats and the need for a short distance cut-off (which, by the way,
naturally appears in String Theory and should be provided by it). 
These results are largely seen to be
independent of the way the cut-off is introduced \cite{lalienafo}. 
For a review see \cite{physrep}.

We believe these works signal the importance of a microcanonical
treatment for string gases even in the free case. In this work, we have set
aside questions related to Jeans instability because we find it compulsory
to make a microcanonical treatment of the problem from first principles in
String Theory paying special attention to its relationship with the work
in \cite{lalienafo} (and references there in). In a completely
independent way, interesting work \cite{barbonrabi} 
has been made on the constraints that 
the classical Jeans instability (as it is gotten from a quantum field theory
in the canonical description) and 
black-hole nucleation have upon the possibility of really having a Hagedorn
"visible" phase that can actually produce black-holes
from highly excited (fat) strings at higher energy. However, no direct 
calculation in the
microcanonical description of the Jeans effect has been done. Knowing as we
do the results recently found about the microcanonical description of classical
self-gravitating systems, we, by no means, find ourselves justified to simply
apply the results about Jeans' instability gotten from a fixed temperature
description.
May be we are being too much conservative, but there are
too many subtleties in the subject that deserve some precaution. Finally, to
understand how Jeans instability appears related to the string loop
expansion computation of the canonical free energy, the work in \cite{eto}
includes a explanation of the relationship between the perturbative loop
expansion and the appearance on Jeans' instability in the 
computation in \cite{gpy}. 

There can be another criticism about the intrinsic instability of the toric box
seen in a cosmological context. Problems like decompactifying
dimensions, the running dilaton, etc. that are characteristic of  toric
zero temperature
vacuum solutions have to be posed, at finite temperature, in a cosmological 
picture where the stress tensor of the string gas enters in the Universe
evolution equations like it was done long time ago in \cite{vs}.
Whether a torus is a good background can then only be seen afterwards although
lots of works have been published and are actively being produced in this
direction (see, as a single example, \cite{brannew}).
Anyhow our main interest in this work is precisely decompactification,
although non-dynamical. Sections 4 and 5 have been devoted 
to show how the toric box can 
decompactify  and we
can reach a situation where, for the string gas, all the dimensions are
effectively open.
   
Let us now focus on what has been the precise subject of this article. 
By computing $\Gamma(E,R,1)$ for a string in compact space (or its
derivative with respect to the energy) by applying a high energy limit
analogous to the one in the regular particle gas, one simply finds a limit
in which the sum upon windings and momenta can be converted to an integral
with big accuracy. This integral is easily seen to be independent of the
volume. After all, converting sums into integrals is a way of getting a more
explicit dependence on the volume than the one seen just in the raw sum. 
Asking why not to use the crude sum instead of the integral has to do with
knowing the volume dependence for the thermodynamical system as it happens
in the classical free particle gas. What is dropped in this approximation,
the phase space boundary contribution, just contains the dependence over the
volume that is not present in the high energy limit. The approximation is
only valid when energy gets big and the radius of the hypertorus is
relatively kept not too big. When the radius grows and so the relation
$\alpha'\,E/R$ is small, the boundary contribution increases its weight and
cannot be dropped, it actually produces a term proportional to the volume as
in the open space case as long as $R$ becomes bigger with fixed energy. In
other words, studying finite size effects in a string system is a subtle
issue because of T-duality.

To better see this, we have shown that what is easily seen true as expressed
in a set of variables is also true when things are re-expressed in a
different set of variables, whenever all the contributions are taken into
account. In the problem at hand, this statement means that the high volume
limit so effortless computed over $\Gamma(E,R,1)$ as written in
\eqref{anchura} can be recovered when the sum over discrete momenta and
windings is represented thorough integrals as though they were continuous
variables. Furthermore, as long as we keep using the representation of the
periodic delta function in terms of phase space coordinates without
introducing more Bernoulli periodic functions, the task of getting the high
volume limit is still a simple one because the integral representing
$\Gamma$ will finally get a contribution from only one of the Dirac delta
functions in the definition of $\delta_{\mathrm{per}}(\vec{n}, \vec{m})$.
Namely, in the limit
$R\longrightarrow
+\infty$ of  the product $\vartheta\lp \rho^{\,\prime\,\2} -
\frac{\vec{n}^{\2}}{R^{\2}} -
\frac{R^2\vec{m}^2}{\a'}\rp\,\delta_{\mathrm{per}}(\vec{n}, \vec{m})$ only
the contribution for $\delta(\vec{m})$ survives. It is now obvious how the high
energy calculations giving an independent of the volume $\Gamma (E, R, 1)$
are related to substituting the sum over windings and momenta in the
computation by a multiple integral. What is now new is the fact that the
volume independent regime can be connected in a smooth way with the regime
proportional to the volume when the equality between the sum and the
integrals expressed thorough the EM formula is taken into account.
Furthermore, there might be intermediate regimes as for the volume
dependence is concerned. As an example, if there really is a regime
proportional to the area of the hypertorus and it is compatible with a
Hagedorn regime in which one single string dominates the system with null
chemical potential, then the many-string system would have an entropy with a
term proportional to the area of the surface enclosing the volume $V$.

Going over to the technical side, there are certain ambiguities when
representing a sum by integrals. The first one is related to the election of
the domain $D$ which contains the lattice points we are summing over. For
example, it holds that $\sum_{k\in \mathbb{Z}}\,\vartheta(l+0.8 -
k)\,\vartheta(k-(n
-0.2))\vartheta(l-n)=\sum_{n}^{l}\,1=\int_{n-0.2}^{l+0.8}\,\dif x-
(B_1(l+0.8)-B_1(n-0.2))$ which certainly equals $l-n+1$ (to simplify,
$n\,,\,l$ are natural numbers, $\langle l.8\rangle=0.8$) and it also equals
$\sum_{k\in[n-0.x\,,\,l+0.y]}\,1$. This shows that there are, in general,
more than one integration domain and that issue affects the value of the
integral over $D$ as the first approximation to the sum. However, this
ambiguity is harmless because it reflects the tolerance from the distance
of a point in the lattice of integer components to the nearest neighbor.
This aspect is negligible in the calculation of
$\Gamma$. What seems more relevant at first sight is the ambiguity in
representing the Kronecker delta function expressing the closed string
constraint. It is immediate to see that the integral
$\int_0^1\,\dif t\,\cos\,2\pi\,t(\Delta -\vec{m}\cdot\vec{n}) =
\displaystyle{\frac{\sin \pi(\Delta - \vec{m}\cdot\vec{n})}{\pi\,(\Delta
-\vec{m}\cdot\vec{n})}}\,\cos\pi(\Delta-\vec{m}\cdot\vec{n})$ is
another representation of $\delta_{\Delta\,,\,\vec{m}\cdot\vec{n}}$ for 
$\vec{m}\,,\,\vec{n}$ integer vectors that could be used instead of the one
in \eqref{ligadura}. It is manifest that, because of the cosine term and
after applying the Euler-Maclaurin summation formula, they are different as
real variable functions although coincide for integer arguments. Taking the
limit $\a'\,E^{\2}\longrightarrow
+\infty$  over the first integral upon $D$ on the right hand side of
\eqref{grande} for both elections  is a good way of estimating the effect
of changing the representation. We get that

\begin{gather}
\lim_{\a'\,E^{\2}\longrightarrow +\infty}\,\a'\,E^{\2}\,
\int_{-1/2}^{+1/2}\,\dif t\,\cos 2\,\pi\,t(\Delta -
\a'\,E^{\2}\,\vec{m}\cdot\vec{n}) =
\delta\lp\frac{\Delta}{\a'\,E^{\2}}-\vec{m}\cdot\vec{n}\rp\nonumber\\
\lim_{\a'\,E^{\2}\longrightarrow +\infty}\,\a'\,E^{\2}\,
\int_{0}^{1}\,\dif t\,\cos 2\,\pi\,t(\Delta -
\a'\,E^{\2}\,\vec{m}\cdot\vec{n}) =
\frac{1}{2}\,\delta\lp\frac{\Delta}{\a'\,E^{\2}}-\vec{m}\cdot\vec{n}\rp
\nonumber
\end{gather}    
where we understand that, when $\Delta\neq 0$, we will have contributions
from terms with $\Delta$ of the order of $\a'\,E^2$. It is clear that after
summing up the contributions from the three integrals in \eqref{grande}, the
result of computing the sum by making integrals is independent of the way we
represent the Kronecker delta.

\section*{Note added}
After disclosing our work, we have found \cite{multi}(and references
therein) in which several multivariate generalizations of the
Euler-Maclaurin formula are presented stressing their usefulness to
numerically compute integrals by sums. It seems clear
that the restrictions this
numerical problem imposes makes this project a more difficult mathematical
task than it was ours.

\section*{Acknowledgments}
We thank Miguel \'Angel Lerma for explaining us some details in relation
with his work on the Bernoulli functions. We thank Igor Sobrado for putting
his computer programmer skills to service our point countings. The work of
M. A. C. gets financial support thorough a fellowship from the F.P.I. Program
of the Spanish MCyT. The work of M. S. is partially financed by a fellowship
from the F.P.U. Program of the Spanish MECD. We all are partially supported
by the Spanish MCyT research projects BFM2000-0357 and BFM2003-00313/FISI.

\section*{Appendix}
\setcounter{equation}{0}
\renewcommand{\theequation}{A.\arabic{equation}}

Let us present the main ingredients of the Euler-Maclaurin formula as
obtained in \cite{lerma1} (see also \cite{lerma2}).

Bernoulli numbers $B_n$ play an important role in the Euler-Maclaurin
formula. Bernoulli numbers are concrete values of the Bernoulli polynomials
($B_n=B_n^*(0)$) that can be defined by the recursive formulas:

\begin{align}
B_{0}^{*}(x) & = 1\\
B_{n}^{*'}(x)& = n\,B_{n-1}^{*}(x)\\
\int_{0}^{1}\,\dif x\,B_{n}^{*}(x)&= 0
\end{align}
with $n\geq 1$

The Bernoulli periodic functions are defined as $B_n(x)=B_n^*(\langle
x\rangle)$, with $\langle x\rangle = x - \lc x\rc =$ fractional part of $x$.
Now M. Lerma in \cite{lerma1} realizes that the odd numbered Bernoulli
numbers are zero except $B_1=-1/2$ and that for all Bernoulli periodic
functions $\int_0^1\,\dif x\,B_n(x)=0$ except for $n=0$ that the integral
gives 1. To make things easier, it is adequate to use a modified version of
the periodic Bernoulli functions that has sense if $B_0$ is a distribution
instead of a function, but that makes the proof of the Euler-Maclaurin
summation formula more natural and, what is important to us, easily
generalizable to multiple integrals and sums.

The Bernoulli periodic distributions are defined thorough the recursive
relations $(n\geq 1)$

\begin{align}
B_0(x) = 1 - \delta_{\mathrm{per}}(x)\\
B_n^{'}(x) = n B_{n-1}(x)\label{derivative}\\
\label{also} \int_0^1\,\dif x\,B_n(x) = 0
\end{align}
Now $B_0(x)$ also verifies \eqref{also}. The differences with the standard
definitions of the Bernoulli functions are in $B_0(x)$ and $B_1(x)$.
$B_1(x)$ is now the function

\be
B_1(x)=\begin{cases}
\langle x\rangle  - \frac{1}{2}&\qquad \text{if $ x \notin \mathbb{Z}$}\\
 0 &\qquad\text{if $ x \in \mathbb{Z}$}
\end{cases}
\ee

This can be checked in relation with \eqref{derivative} by computing
$B_1(x) = \int_0^x\,\dif y\, B_0(y)$. It gives

\be\begin{split}
B_1(x) = &\, x - \sum_{k\in\mathbb{Z}}\,
\int_0^x\,\dif y\quad\delta\lp y - k\rp
\\  = &\, x - \sum_{k\in\mathbb{Z}}\,\lc\vartheta(x-k) - \vartheta(-k)\rc=
\begin{cases}
\langle x\rangle -\frac{1}{2}&\qquad 
\text{if $x\notin\mathbb{Z}$}\\
0 &\qquad 
\text{if $x\in\mathbb{Z}$}
\end{cases}\end{split}
\ee
where it is worth to remark that the Dirac delta has been integrated
using $\delta\lp x\rp = \frac{\dif}{\dif x}\,\,\vartheta(x)$ with
$\vartheta(x)$ the Heaviside function of value $\frac{1}{2}$ for $x=0$.

The Fourier series representation of the Bernoulli functions will be
exceedingly useful when generalizing Lerma's proof to multivariate sums.
We have

\be
B_n(x) = -\frac{n!}{(2\pi\mathrm{i})^n}\,\sum_{\substack{k=-\infty\\k\neq
0}}^{+\infty}\,\frac{\mathrm{e}^{\,2\pi\mathrm{i}kx}}{k^{\,n}}
\ee

The proof of the EM summation formula is now very easy and starts
with 
$\int_a^b\dif x f(x) - \sum_{a\leq k\leq b}^{'}\, f(k) = 
\int_a^b\,\dif x \,B_0(x)$. This is not a tautology because we can use that 
$B_1^{'}(x) = B_0(x)$ to integrate by parts on the right hand side. This can
be done successively by using \eqref{derivative}. The prime over the sum
results again from the fact that $\vartheta(0)=1/2$. And the summation
formula emerges beautifully.

To generalize this to multiple sums is straightforward. We can easily
write down the periodic Dirac delta function of $n$ variables (see
\eqref{periodic}) and define the rest of the multivariate Bernoulli periodic
functions thorough their Fourier expansions as two series of scalar and
vectorial functions.

\begin{gather}
B_n(\vec{x}) =
-\frac{n!}{(2\pi\mathrm{i})^n}\,
\sum_{\substack{\vec{k}\in\mathbb{Z}\\\vec{k}\neq\vec{0}}}
\frac{\mathrm{e}^{\,2\pi\mathrm{i}\vec{k}\cdot\vec{x}}}{\lvert\vec{k}\rvert^n}
\quad\text{if $n\in (2\mathbb{Z})$}
\\
\vec{B}_n(\vec{x}) =
-\frac{n!}{(2\pi\mathrm{i})^n}\,
\sum_{\substack{\vec{k}\in\mathbb{Z}\\\vec{k}\neq\vec{0}}}
\frac{\mathrm{e}^{\,2\pi\mathrm{i}\vec{k}\cdot\vec{x}}}
{\lvert\vec{k}\rvert^{n+1}}
\,\,\vec{k}\quad 
\text{if $n\in (2\mathbb{Z} + 1)$}
\end{gather}
Now $1-B_0(\vec{x})= \sum_{\vec{k}}\,\delta(\vec{x}-\vec{k})$, and 
the recurrence relations are $\vec{\nabla}\cdot
\vec{B}_{2n+1}(\vec{x})=(2n+1)\,B_{2n}(\vec{x})$ and
$\vec{\nabla}\,B_{2n+2}(\vec{x})=
(2n+2)\,\vec{B}_{2n+1}(\vec{x})$ with $n\geq 0$.

The expression we get for a sum over integer vectors
belonging to a compact region $D$ defined as a subset of $\mathbb{R}^n$, 
with $\vec{B}_1(\vec{x})$ and $B_2(\vec{x})$ explicitly shown, is

\be
\left.\sum_{\vec{k}\in
D}\right.^{'}\,g(\vec{k}) =
\int_{D}\dif\vec{x}\,\,g(\vec{x})
-\int_{\partial D}\,\, \dif\vec{S}
\cdot \lp g\lp\vec{x}\rp\vec{B}_1\rp +
\frac{1}{2}\,\int_{\partial D}\,\dif \vec{S}\cdot\,\lp B_2\,\vec{\nabla}g\rp
- \frac{1}{2}\,\int_{D}\,\dif\vec{x}\,B_2\,\Delta\,g
\ee
 
The prime means that the lattice points $\vec{k}_0$ that are just over the
surface boundary will contribute a factor (different from one in general)
times $g(\vec{k}_0)$. In the case the domain $D$ is defined as the region
bounded by the surface defined by the subset of $\mathbb{R}^n$ of points
$\vec{x}$ such that  $f(\vec{x})=0$, a single Heaviside function will serve
to write $ 
\sum_{\vec{k} \in
D}\mbox{}^{'}\,g(\vec{k}) = \sum_{\vec{k}\in \left.\mathbb{Z}\right.^n}\,
g(\vec{k})\,
\vartheta(f(\vec{k}))$, so the prime will mean that the integer vectors
just on the boundary surface, $\vec{k_0}$, will contribute $g(\vec{k_0})/2$
because the value of the step function is 1/2 when its argument gets null.

In our applications, we have found a sum  in which the domain of integration
is one that cannot be defined using a single step function, but a
combination of products of several of them. More precisely, it is the sum
over all the nine dimensional integer
$\vec{k}$-vectors with the exception of the vector $\vec{0}$. We have
chosen to take out the origin bounding it by a cube of two length side
centered at the origin. The domain $D$ will then be the region outside the
cube. To get only a contribution to the sum from points in $D$, we must
insert in the sum a factor

\be
\sum_{i_1\neq i_2\neq ...\neq i_9}\,\sum_{k=1}^9\,\,
\frac{1}{k!}\,\frac{1}{(9-k)!}\,
\vartheta\lp x_{i_1}^{\2} -1\rp\,...\,\vartheta\lp x_{i_k}^{\2}-1\rp
\vartheta(1- x_{i_{k+1}}^{\2})\,...\,\vartheta\lp 1-x_{i_9}^{\2}\rp  
\ee

From here it is easy to see that, for example, a vertex of the cube like
\linebreak[4] $(1,1,1,1,1,1,1,1,1)$  would contribute
\be
\sum_{k=1}^9\,\frac{1}{2^{\,9}}\,\binom{9}{k} = 1-\frac{1}{2^{\,9}}
\ee 
If one wants to compute the sum without any restriction, then one has to sum
$1/2^{\,9}$ of  $g\lp(1,1,1,1,1,1,1,1,1)\rp$ to the integrals giving the
primed sum.

In our case, we have that $g(\vec{k})$ is constant so one can readily show that
the integral faithfully represents the primed sum because the integral over the
surface of the cube vanishes.

\end{document}